\documentclass[preprint,aps,showpacs,draft]{revtex4}

\newcommand{\sr}{\scriptstyle}

\usepackage{graphicx}
\usepackage{dcolumn}
\usepackage{amsmath}

\begin{document}


\title{Director configurations
of a nematic liquid crystal confined in a toroidal geometry.
A finite element study.}

\author{Joachim Stelzer}
 \email{js69190@gmx.de} 
\affiliation{Schwetzinger Str.~20, 69190 Walldorf (Baden), Germany}

\author{Ralf Bernhard}
\affiliation{IMPACT Messtechnik GmbH, 71332 Waiblingen, Germany}



\begin{abstract}
Various director configurations of a nematic liquid crystal, confined
in a toroidal volume and subject to an external magnetic field, were
evaluated numerically by performing finite element calculations in
three dimensions. The equilibrium director field is found from a
minimization of the elasto-magnetic energy of the nematic. The
director at the inner surface of the torus is fixed, either along
homeotropic or tangential direction. We consider both a homogeneous
and an azimuthal magnetic field, with its direction being in conflict
with the respective surface anchoring.

The relative stability of topologically inequivalent configurations is
investigated in dependence on the strength of the magnetic field and
the geometrical parameters of the confining torus. We find that in a
toroidal geometry the director ``escape'' from a disclination, which
usually occurs in a cylindrical tube, is absolutely stable only for
weak magnetic fields and tangential anchoring, whereas for homeotropic
anchoring director fields containing disclination rings are more
favorable. When a strong homogeneous magnetic field is exerted
laterally, the resulting director field is not any more axially
symmetric, instead, it reveals curved disclinations of finite length.
\end{abstract}


\maketitle


\section{INTRODUCTION}

Since the discovery of polymer-dispersed liquid crystals (PDLC)
\cite{Fergason,Doane,Drzaic} the confinement of nematics in curved
geometries attracts considerable interest \cite{Crawford}.  Small
nematic droplets embedded in a polymer matrix give rise to light
scattering effects \cite{Zumer,Montgomery,Whitehead} which are
controllable by an external electric field and thus can be applied in
electro-optical light shutters of high contrast.  On the other hand,
from the viewpoint of basic research, nematics confined by curved
surfaces are challenging as they exhibit non-trivial director fields.
Due to the large surface-to-volume ratio the boundary conditions at
the surface strongly influence the emerging structures which reveal a
large variety of topological defects \cite{Crawford}.  The structure
of nematics confined in a spherical droplet or a cylindrical capillary
tube has been studied in detail. For the spherical droplet,
homeotropic surface anchoring mainly leads to a radial point defect in
the center of the droplet \cite{Williams,Saupe,Meyer,Lavrentovich} or,
alternatively, to a defect ring in the equatorial plane of the droplet
\cite{Mori,Schopohl1,Terentyev,Sonnet,Gartland1,Kralj}.  Tangential
anchoring creates two surface point defects, denoted boojums, at
opposite poles of the droplet surface \cite{Candau,Kurik}. (For some
years, the inverse problem of a nematic surrounding an isotropic
spherical droplet has been addressed, too
\cite{Terentyev,Kuksenok,Ruhwandl,Poulin,Lubensky,Stark1}.)  On the
other hand, in a cylindrical geometry, homeotropic or tangential surface
anchoring does not lead to a line defect, instead, this disclination
generally turns out to be unstable against an ``escape'' of the
director field lines along the symmetry axis of the tube
\cite{Cladis,Williams,Meyer,Lyuksyutov,Schopohl2}.

In our contribution, we consider an advanced scenario, namely, the
confinement of a nematic liquid crystal in a torus.  This is an
elementary paradigm for a nematic filling a volume which is not simply
connected. For some years now, the method of finite elements has been
applied to the study of liquid crystals and polymers in complex
geometries \cite{Gartland2,Stark2,Feng}. Accordingly, we performed
finite element calculations in three dimensions for a toroidal
geometry, thus enabling the investigation of fairly complicated
situations.  Both homeotropic and tangential boundary conditions for
the director field at the torus surface are taken into account.  In
addition, an external magnetic field is exerted on the nematic whose
direction is in conflict with the boundary conditions imposed.  By
minimizing the elasto-magnetic free energy of the nematic the
equilibrium director configurations are found. We discuss the relative
stability of various topological defect structures in dependence of
the strength of the magnetic field and the geometrical parameters of
the confining torus.

The organization of the article is as follows. In Section II
the finite element method for director fields in a torus
is introduced. Section III presents a survey of equilibrium
director configurations for various situations.
Finally, Section IV contains some concluding remarks.

\section{TOROIDAL GEOMETRY AND FINITE ELEMENT METHOD}

The geometry of our system is confined by a torus with its symmetry
axis along the $z$ direction and its large circle in the $x$-$y$
plane. The radii of the large and small circle are $A$ and $B$,
respectively (Figure 1). The torus is filled by a nematic liquid
crystal whose director field is fixed at the inner surface of the
torus.  According to the toroidal geometry there are three distinct
boundary conditions of Dirichlet type. Their respective directors at
the surface on sections of the torus across the $x$-$z$ and $y$-$z$
coordinate planes are visualized from an oblique view in Figure 2.
(Note that for sake of clarity not all directors that were calculated
on the grid are displayed.)  The director can be anchored either
perpendicular ({\em homeotropic}) or tangential to the surface.
Tangential anchoring can be established along sections across the
small circle of the torus ({\em transversal-tangential}) or,
alternatively, following the direction of its large circle ({\em
azimuthal-tangential}).  In addition to the boundary conditions at the
surface, the director field is affected by an external magnetic field.
We examine the influence of both a homogeneous magnetic field and an
inhomogeneous magnetic field of axial symmetry.  The latter is created
by an electric current through a wire of infinite length along the
symmetry axis of the torus, giving rise to closed magnetic field lines
along the azimuthal direction. Whereas the azimuthal magnetic field as
well as the homogeneous field along the $z$ direction still preserve
axial symmetry, this rotational symmetry does not hold any more for a
``lateral'' homogeneous field along the $x$ axis.  Obviously, for such
a scenario, a finite element calculation in three dimensions is
necessary.

The particular combinations of boundary conditions and magnetic fields
that we considered in our calculations will be introduced in the
following section. For each combination the equilibrium director
fields are found from a numerical minimization of the free energy $F$
of the nematic, with the corresponding energy density
\begin{equation}
\label{energydens1}
{\cal F} = \frac{\sr 1}{\sr 2}\,K_{11}\,
({\mathrm{div}}\,{\bf n})^2
+ \frac{\sr 1}{\sr 2}\,K_{22}\,
({\bf n}\cdot{\mathrm{curl}}\,{\bf n})^2
+ \frac{\sr 1}{\sr 2}\,K_{33}\,
({\bf n}\times{\mathrm{curl}}\,{\bf n})^2
- \frac{\sr 1}{\sr 2}\,\mu_0\,\Delta\mu\,({\bf n}\cdot{\bf H})^2.
\end{equation}
The first three terms in (\ref{energydens1}) form the elastic free
energy density due to {\em splay}, {\em twist} and {\em bend}
deformations, respectively, of the director field ${\bf n}$, according
to Oseen, Z\"ocher and Frank \cite{Frank}.  $K_{11}$, $K_{22}$ and
$K_{33}$ are the elastic constants for the three deformation modes
mentioned above.  The last expression in (\ref{energydens1})
represents the coupling of the director field to an external magnetic
field ${\bf H}$.  Here the anisotropy of the magnetic permeability
$\Delta\mu$ of the nematic acts as the coupling strength. ($\mu_0$ is
the magnetic field constant.)

For numerical treatment, the nematic free energy density
(\ref{energydens1}) is expressed in a dimensionless form, {\em i.e.},
in units of $K_{33}/a^2$, where $a$ is a measure for the
characteristic length scale of our system (usually in the range of a
micron).  The director field is determined by the {\em tilt} and {\em
twist} angle, $\Theta$ and $\Phi$, respectively,
\begin{equation}
\label{director}
{\bf n} = \hat{\bf{x}}\,\sin\Theta\,\cos\Phi
+ \hat{\bf{y}}\,\sin\Theta\,\sin\Phi
+ \hat{\bf{z}}\,\cos\Theta.
\end{equation}
Therefore, the reduced free energy $\overline{F}$, {\em i.e.}, the
volume integral of the free energy density becomes a functional of the
tilt and twist angle,
\begin{eqnarray}
\label{energydens2}
\overline{F}[\Theta,\Phi] &=& 
\int\,\mbox{d}^3 \overline{x}\,\left\{
\frac{\sr 1}{\sr 2}\,\overline{K}_{11}\,
\left[
(\partial_x\Theta\,\cos\Phi + \partial_y\Theta\,\sin\Phi)\,\cos\Theta
- \partial_z\Theta\,\sin\Theta \right. \right.
\\
& & 
- \left. (\partial_x\Phi\,\sin\Phi 
- \partial_y\Phi\,\cos\Phi)\,\sin\Theta \right]^2
\nonumber \\ 
& & + \frac{\sr 1}{\sr 2}\,\overline{K}_{22}\,
\left[
\partial_x\Theta\,\sin\Phi - \partial_y\Theta\,\cos\Phi \right.
\nonumber \\
& & 
\left. + (\partial_x\Phi\,\cos\Phi 
+ \partial_y\Phi\,\sin\Phi)\,\sin\Theta\,\cos\Theta 
- \partial_z\Phi\,\sin^2\Theta \right]^2
\nonumber \\
& & + \frac{\sr 1}{\sr 2}\,
\left( \left[ (\partial_x\Theta\,\cos\Phi 
+  \partial_y\Theta\,\sin\Phi)\,\cos\Phi\,
\sin\Theta\,\cos\Theta
+ \partial_z\Theta\,\cos^2\Theta\,\cos\Phi \right. \right.
\nonumber \\ 
& & 
\left. - (\partial_x\Phi\,\cos\Phi +  \partial_y\Phi\,\sin\Phi)\,\sin\Phi\,
\sin^2\Theta
- \partial_z\Phi\,\sin\Theta\,\cos\Theta\,\sin\Phi \right]^2
\nonumber \\ 
& & + \left[ (\partial_x\Theta\,\cos\Phi 
+  \partial_y\Theta\,\sin\Phi)\,\sin\Phi\,
\sin\Theta\,\cos\Theta
+ \partial_z\Theta\,\cos^2\Theta\,\sin\Phi \right. 
\nonumber \\ 
& & 
\left. + (\partial_x\Phi\,\cos\Phi +  \partial_y\Phi\,\sin\Phi)\,\cos\Phi\,
\sin^2\Theta
+ \partial_z\Phi\,\sin\Theta\,\cos\Theta\,\cos\Phi \right]^2
\nonumber \\ 
& & \left. 
+ \left[ (\partial_x\Theta\,\cos\Phi 
+  \partial_y\Theta\,\sin\Phi)\,\sin^2\Theta
+ \partial_z\Theta\,\sin\Theta\,\cos\Theta \right]^2 \right) 
\nonumber \\ 
& & \left.
- \frac{1}{2\overline{\xi}}\,
\left[ h_x\,\sin\Theta\,\cos\Phi + h_y\,\sin\Theta\,\sin\Phi
+ h_z\,\cos\Theta \right]^2 \right\}\; .
\nonumber
\end{eqnarray}
In (\ref{energydens2}), all lengths are in reduced units of $a$.  The
reduced elastic constants are $\overline{K}_{11} = K_{11}/K_{33}$ and
$\overline{K}_{22} = K_{22}/K_{33}$.  $\overline{\xi} = \xi/a$ is the
reduced magnetic coherence length, with $\xi =
\sqrt{K_{33}/(\mu_0\Delta\mu)}\cdot (1/H)$. $H$ is the strength of the
magnetic field ${\bf H}$ and $(h_x, h_y, h_z)$ are the Cartesian
components of the local unit vector along the direction of ${\bf
H}$. For the homogeneous magnetic field this unit vector is constant
along the $z$ or $x$ direction, whereas for the azimuthal field it
becomes $( -y/\rho,\,x/\rho,\,0)$, with cylindrical coordinate $\rho =
\sqrt{x^2 + y^2}$.

The parametrization of our calculations is based on the nematic liquid
crystal pentylcyanobiphenyl (5CB) \cite{Handbook}.  Its elastic
constants are $K_{11} = 4.2\times 10^{-12}\,$N, $K_{22} = 2.3\times
10^{-12}\,$N, $K_{33} = 5.3\times 10^{-12}\,$N, which amounts in the
ratios $\overline{K}_{11} = 0.79$ and $\overline{K}_{22} = 0.43$.  The
strength of the magnetic field is varied from zero up to $1.5\times
10^7\,$A/m which, with a magnetic anisotropy of $1.2\times 10^{-7}$,
means a reduced inverse coherence length in the range between zero and
2.5. Besides the magnetic coherence length, the geometrical parameters
of the confining torus are changed.  Its large radius ranges from 
$A = 7$ to 9, whereas its small radius is between $B = 2.0$ and 2.4.

The free energy functional (\ref{energydens2}) is now discretized on
an irregular lattice consisting of between 18744 and 26624
tetrahedron-like finite elements (depending on system size), which
corresponds to 4892 up to 6848 vertices or, equivalently, to an
average lattice constant of about 0.48. The tilt and twist angles are
then defined on the vertices of the finite element grid and the total
free energy is obtained numerically from an integration according to
(\ref{energydens2}).  (For further technical details on the finite
element method for nematics we refer to the Appendix.) Starting from
suitable initial configurations, the equilibrium director fields that
correspond to local minima of the total free energy are calculated via
a standard Newton-Gau{\ss}-Seidel relaxation method, where the
director angles $\Theta_i$ and $\Phi_i$ on each internal vertex $i$
are immediately replaced by their respective corrected values,
according to
\begin{eqnarray}
\label{Seidel1}
\Theta_i^{\mathrm{new}} &=&
\Theta_i^{\mathrm{old}} 
- \frac{\partial F/\partial \Theta_i}{
\partial^2 F/\partial \Theta_i^2} \enspace ,
\\
\label{Seidel2}
\Phi_i^{\mathrm{new}} &=&
\Phi_i^{\mathrm{old}} 
- \frac{\partial F/\partial \Phi_i}{
\partial^2 F/\partial \Phi_i^2}\; .
\end{eqnarray}
The functional derivatives in (\ref{Seidel1}) and (\ref{Seidel2}) are
again evaluated numerically.

\section{DIRECTOR CONFIGURATIONS IN A TORUS}

Let us first introduce the different combinations of boundary and
initial conditions as well as the external magnetic field.  Both
for homeotropic boundary conditions and for transversal-tangential
anchoring, the following combinations of initial configurations and
magnetic field direction are chosen: $(a)$ a $+1$ ring defect with a
homogeneous magnetic field along the $z$ axis, $(b)$ both director and
magnetic field homogeneous along $z$, $(c)$ both director and magnetic
field in azimuthal direction.  Whereas $(a)$ amounts in a distortion
of the director field, $(b)$ yields a topologically distinct
configuration, consisting of two $+ \frac{1}{2}$ ring defects.  $(c)$
favors the ``escape'' of the director field along the azimuthal
direction. For azimuthal-tangential anchoring we assume initial
configurations with director and magnetic field along the $z$ and
azimuthal direction, respectively.  Finally, we consider a
``laterally'' homogeneous director and magnetic field, {\em i.e.},
along the $x$ direction, as initial condition for several types
of anchoring.

We start the discussion on director configurations in a toroidal
volume by considering homeotropic surface anchoring. With the external
magnetic field being switched off, there are mainly three possible
solutions. Their visualization in the sections of the torus across the
$x$-$z$ and $y$-$z$ cooodinate planes is presented in Figure 3.  If
the director field does not exhibit an azimuthal component, there are
closed disclinations, namely, defect rings. These rings can be
classified according to the strength of the disclination
\cite{Garel,Nakanishi}. In our toroidal system we expect either one
ring of strength $+1$ located in the central region of the torus
(Fig.~3a) or, alternatively, two rings of strength $+\frac{1}{2}$ at
its outer parts (Fig.~3b).  Another possible director configuration is
the ``escape'' structure (Fig.~3c) where the director in the bulk
turns towards the azimuthal direction. In this way, the occurence of
defect rings would be suppressed.

For homeotropic anchoring, both the homogeneous magnetic field along
$z$ and the azimuthal magnetic field are in conflict with the director
orientation at the torus surface. The homogeneous field is expected to
yield a ring configuration, whereas the azimuthal field favors the
escape configuration. Figure 4 displays the energy of the three
director configurations versus the inverse magnetic coherence length
which, as indicated above, is proportional to the strength of the
magnetic field. Please note that each of the three curves is related
to the type of magnetic field favoring the respective bulk director
orientation (see Figure caption for details).  In order to compare
systems of different size (the torus radii are $A=7$, $B=2$ and $A=7$,
$B=2.4$ in the top and bottom part, respectively, of Figure 4) the
total energy is divided by the torus volume
$4\pi^2\,A\,B^2$. Surprisingly, with zero magnetic field the escape
structure is not stable.  This is contrary to the director field in a
cylindrical tube where disclinations are usually avoided by a director
escape along the axis of the cylinder
\cite{Cladis,Williams,Meyer,Lyuksyutov,Schopohl2}.  In a torus,
however, the escape seems to be suppressed. Therefore, we attribute
this behavior to the additional curvature along the azimuthal
direction which is absent in a cylinder.  Even when the small circle
of the torus is increasing, the escape is far from being stable
(Fig.~4, bottom). It can be stabilized against the global minimum only
by a strong magnetic field along the azimuthal direction which, for
our parametrization, should correspond to a coherence length of at
least 0.65.

From Figure 4 it is revealed that the global minimum configuration is
the two-ring director field (Fig.~3b).  Although the total topological
charge is the same as in the one-ring configuration, it is
energetically preferred, because a major percentage of the toroidal
volume is filled by an almost homogeneous director field along
$z$. (Obviously, the relative stability of the two-ring configuration
is further enhanced by a homogeneous magnetic field along $z$.) Let us
compare the situation to a spherical nematic drop subject to
homeotropic surface anchoring. There, the analoga of the one-ring and
two-ring configuration are a radial point defect
\cite{Williams,Saupe,Meyer,Lavrentovich} and an equatorial defect ring
(``Saturn ring'')
\cite{Mori,Schopohl1,Terentyev,Sonnet,Gartland1,Kralj}.  When the
radius of the drop is not too small, the point defect is always
favored to the ring, due to its much lower energy. In the torus,
however, the two competing defect configurations are much more
similar, and therefore it is the smooth director field rather than the
defects which decides about their relative stability.

The situation is qualitatively changed for transversal-tangential
surface anchoring. Again the one-ring, two-ring and escape
configurations are possible solutions which are visualized in Figure
5. The one-ring configuration is known as closed vortex line
(Fig.~5a), whereas the defects in the two-ring director field are
located on the very surface of the torus (Fig.~5b). The escape
configuration is analogous to the one for homeotropic anchoring
(Fig.~5c). Interestingly, for not too large radii of the small circle
of the torus and for weak magnetic fields, the escape is now
absolutely stable (Fig.~6, top).  A homogeneous magnetic field along
$z$ with coherence length larger than 0.35 is needed to stabilize the
closed vortex or surface defects against the escape. On the other
hand, if we increase the small radius of the torus, the escape is
suppressed again (Fig.~6, bottom).  The stability of the escape for
transversal-tangential anchoring can be understood from the elastic
properties of the nematic sample.  As stated, our parametrization is
based on 5CB, which is an example for a rod-like molecule with elastic
anisotropy $K_{33}> K_{11} > K_{22}$.  The escape configuration with
transversal-tangential anchoring mainly requires {\em twist}
deformations, whereas for the escape with homeotropic anchoring 
{\em bend} deformations are predominant.  For rod-like molecules, $K_{22}$
usually is the smallest elastic constant. Undergoing an escape, the
increase of the elastic energy due to {\em twist} deformations is not
too large, and it can be overcompensated by the disappearance of the
defect rings.  (For disc-like molecules the elastic anisotropy is
reversed, which should cause a stable escape configuration in case of
homeotropic rather than tangential anchoring.)

As mentioned above, for increasing small radius of the torus, the
escape becomes unstable against the two-ring configuration.  The
latter enables an almost homogeneous director field in the bulk, apart
from the two closed surface disclinations.  Here, the two-ring
director field is the analogon of the so-called ``boojum''
configuration in a spherical nematic drop with tangential anchoring,
where two point defects occur at opposite poles of the sphere
\cite{Candau,Kurik}.

Whereas for homeotropic and transversal-tangential anchoring we find a
smooth dependence of the free energy on the strength of the magnetic
field, the situation is different for azimuthal-tangential
anchoring. Obviously, an azimuthal director field now is the stable
configuration for zero magnetic field (Fig.~7a).  When applying a
homogeneous magnetic field along $z$ (Fig.~7b), we find a threshold
field strength which must be exceeded in order to change the director
configuration (Fig.~8). For magnetic fields slightly above the
threshold only the director in the central part of the torus turns
into $z$ direction, which means an abrupt jump of the energy to larger
values. When the magnetic field further increases, the director in the
outer regions of the torus gradually aligns parallel to the magnetic
field and, therefore, the energy is again decreasing. As we notice
from Figure 8, the threshold field crucially depends on the small
radius of the torus: for larger $B$ the onset of the reorientation
already occurs at smaller correlation lengths, mainly because the
elastic deformation can be spread over a larger region. On the other
hand, the large radius $A$ of the torus does not significantly
influence the emerging director fields.

Finally, let us turn towards a more complicated scenario. When a
strong homogeneous magnetic field is applied along the $x$ direction,
the system does not possess axial symmetry any more. Therefore, we
have to analyze the three-dimensional visualizations in detail.  We
first consider a director configuration with homeotropic anchoring,
subject to a lateral magnetic field of reduced coherence length
$\overline{\xi} = 2.5$. This field is strong enough to force the
director in the bulk to align along $x$, which can clearly be seen
from a top view onto the section of the torus across the $x$-$y$ plane
(Fig.~9, top). When we look at the sections across the remaining
coordinate planes (Fig.~9, bottom), two limits can be discerned in the
director alignment. Due to the director orientation in the bulk along
$x$, locally in the $y$-$z$ plane there is an escape, comparable to
Figure 3c. On the other hand, the director configuration in the
$x$-$z$ plane strongly resembles the two-ring configuration of
Figure 3b, but with the disclinations now appearing close to the upper
and lower edge of the torus rather than in the $x$-$y$ plane.
Therefore, the fact that the escape configuration does not exhibit any
defects means that the disclinations occuring in the $x$-$z$ plane are
curved, but not closed. Instead, they are of finite length,
terminating at points where the {\em bend} deformation that causes the
escape structure becomes more favorable energetically. Thereby, the
extension of the disclinations covers an azimuthal range of
$\Delta\phi \approx \pi/3$, as revealed from a more thorough analysis
of the configuration.

An analogous director reorientation is observed for
transversal-tangential anchoring (Fig.~10). Again, the escape
structure occurs in the $y$-$z$ plane, whereas in the $x$-$z$ plane we
now find the two-ring surface defects according to Fig.~5b. Here,
these surface defects appear in the $x$-$y$ plane (Fig.~10,
bottom). Due to the smaller elastic {\em splay} energy, which
dominates the reorientation in case of tangential anchoring, the
escape covers a larger azimuthal range, thus reducing the extension of
the surface defects to $\Delta\phi \approx \pi/4$.

\section{REMARKS}

\begin{enumerate}

\item Summarizing our work, we have performed finite element
calculations for director configurations of a nematic liquid crystal
confined in a torus. Three different types of fixed surface anchoring,
according to the toroidal geometry, were investigated.  In addition,
an external magnetic field was applied along directions forcing a
reorientation of the director in the bulk. With zero magnetic field,
the director ``escape'' along azimuthal direction turns out to be
stable only for transversal-tangential anchoring. However, it is
completely suppressed for homeotropic surface orientation where a
configuration containing two defect rings is favored. For
transversal-tangential anchoring these ring disclinations are located
on the surface of the torus. When the surface anchoring is along the
azimuthal direction, for a magntic field applied along the symmetry
axis of the torus, there is a threshold value which must be exceeded
in order to force an abrupt reorientation of the bulk director.

\item Our three-dimensional finite element study allows the
investigation of director configurations which are not any more
axially symmetric. As an example we chose a scenario where a strong
homogeneous magnetic field is applied laterally to the torus.  The
emerging director configuration turns out to contain both the
``escape'' along azimuthal direction and the disclinations, depending
on the local azimuthal direction relative to the external magnetic
field.  The disclinations are curved, but of finite length,
terminating at the intermediate regions where the energy gain due to
the ``escape'' structure suppresses the formation of defects. Thereby,
the extension of the disclinations depends on the type of surface
anchoring.

\item The system of a nematic confined in a torus is the prototype of
a complex geometry which, unlike nematic drops, is not any more simple
connected. Finite element calculations in three dimensions is the
appropriate technique to determine the equilibrium director
configurations in such a system. Therefore, the results obtained can
serve as a starting point for the investigation of more advanced
scenarios, {\em e.g.}, deformations of the confining torus or the
presence of isotropic droplets immersed in the nematic.

\end{enumerate}

\appendix

\section{FINITE ELEMENT TECHNIQUE FOR NEMATICS IN THREE DIMENSIONS}

The method of finite elements is a means to minimize a functional
defined on an arbitrarily complex geometry. In our case, this
functional is the reduced elasto-magnetic free energy of a nematic
liquid crystal (\ref{energydens2}), depending on the {\em tilt} and
{\em twist} angle fields. We start by subdividing the toroidal volume
into $L$ small tetrahedron-like finite elements. The total free energy
is then expressed as the sum over the energy content of the $L$ finite
elements, the director angles now being defined on the $N$ vertices of
the grid. Each finite element has four corner vertices, thereby, the
relation between elements and vertices is established by a neighbor
list. Alltogether, this yields the discretized version of the reduced
free energy (\ref{energydens2}),
\begin{equation}
\label{energydiscr}
\overline{F}[\Theta,\Phi] 
= \sum_{\alpha=1}^{L}\,\overline{F}_{\alpha}\left[\Theta_1^{(\alpha)},
\Theta_2^{(\alpha)}, \Theta_3^{(\alpha)}, \Theta_4^{(\alpha)},
\Phi_1^{(\alpha)}, \Phi_2^{(\alpha)}, \Phi_3^{(\alpha)}, 
\Phi_4^{(\alpha)}\right].
\end{equation}
In (\ref{energydiscr}), the sum over $\alpha$ runs over all $L$ finite
elements, $\overline{F}_{\alpha}$ is the reduced free energy of the
$\alpha$-th element. $\Theta_{j}^{(\alpha)}$ and $\Phi_{j}^{(\alpha)}$
are the director angles on the $j$-th corner vertex ($j = 1,2,3,4$) of
the $\alpha$-th element. For calculating $F_{\alpha}$, we introduce
the Oseen-Z\"ocher-Frank free energy density $\overline{\cal F}$,
according to (\ref{energydens2}),
\begin{equation}
\label{energyfe}
\overline{F}_{\alpha} = \int_{V_\alpha}\,
\mbox{d}^3 \overline{x}\,\overline{\cal F}
\left[\Theta({\bf r}), \Phi({\bf r})\right].
\end{equation}
Here, the integral is over the volume of the respective finite
element.  In order to unify the treatment of the finite elements, we
apply a coordinate transformation from the vertex coordinates 
$x$, $y$, $z$ to {\em natural} coordinates $u$, $v$, $w$,
\begin{eqnarray}
x^{(\alpha)} &=& x_1^{(\alpha)}
+ \left( x_2^{(\alpha)} - x_1^{(\alpha)}\right)\,u
+ \left( x_3^{(\alpha)} - x_1^{(\alpha)}\right)\,v
+ \left( x_4^{(\alpha)} - x_1^{(\alpha)}\right)\,w \enspace ,
\\
y^{(\alpha)} &=& y_1^{(\alpha)}
+ \left( y_2^{(\alpha)} - y_1^{(\alpha)}\right)\,u
+ \left( y_3^{(\alpha)} - y_1^{(\alpha)}\right)\,v
+ \left( y_4^{(\alpha)} - y_1^{(\alpha)}\right)\,w \enspace ,
\\
z^{(\alpha)} &=& z_1^{(\alpha)}
+ \left( z_2^{(\alpha)} - z_1^{(\alpha)}\right)\,u
+ \left( z_3^{(\alpha)} - z_1^{(\alpha)}\right)\,v
+ \left( z_4^{(\alpha)} - z_1^{(\alpha)}\right)\,w\; .
\end{eqnarray}
In the transformations above, $x_j^{(\alpha)}$, $x_j^{(\alpha)}$,
$x_j^{(\alpha)}$ ($j =1,2,3,4$) are the cartesian coordinates of the
$j$-th corner vertex of the $\alpha$-th finite element.  In the new
coordinates all finite elements are congruent. The free energy
$\overline{F}_\alpha$ now becomes
\begin{equation}
\label{energynat}
\overline{F}_{\alpha} = \Delta_{\alpha}\,\int_{0}^{1}\,\mbox{d}w\,
\int_{0}^{1-w}\,\mbox{d}v\,\int_{0}^{1-v-w}\,\mbox{d}u\,
\overline{\cal F}[\Theta(u,v),\Phi(u,v)] \enspace ,
\end{equation}
with the Jacobian determinant $\Delta_{\alpha}$ of the coordinate
transformation.
When the grid is sufficiently fine, the director angle fields within
one finite element can be approximated by a linear interpolation over
the volume of the element,
\begin{eqnarray}
\label{interpol1}
\Theta^{(\alpha)} &=& \Theta_1^{(\alpha)}
+ \left( \Theta_2^{(\alpha)} - \Theta_1^{(\alpha)}\right)\,u
+ \left( \Theta_3^{(\alpha)} - \Theta_1^{(\alpha)}\right)\,v
+ \left( \Theta_4^{(\alpha)} - \Theta_1^{(\alpha)}\right)\,w \enspace ,
\\
\label{interpol2}
\Phi^{(\alpha)} &=& \Phi_1^{(\alpha)}
+ \left( \Phi_2^{(\alpha)} - \Phi_1^{(\alpha)}\right)\,u
+ \left( \Phi_3^{(\alpha)} - \Phi_1^{(\alpha)}\right)\,v
+ \left( \Phi_4^{(\alpha)} - \Phi_1^{(\alpha)}\right)\,w\; .
\end{eqnarray}
After inserting (\ref{interpol1}) and (\ref{interpol2}) into
(\ref{energynat}), due to the small size of the finite elements the
integrand in (\ref{energynat}) is evaluated in the center of the
finite element only. The integrals in (\ref{energynat}) then yield the
product of the reduced free energy density 
$\overline{\cal F}[\Theta^{(\alpha)},\Phi^{(\alpha)}]$ 
and the volume of the finite element,
\begin{equation}
\overline{F}_{\alpha} = \frac{\sr 1}{\sr 6}\,\Delta_{\alpha}\,
\overline{\cal F}[\Theta^{(\alpha)},\Phi^{(\alpha)}].
\end{equation}
In order to calculate the reduced free energy density ${\cal F}$, the
director angles entering (\ref{energydens2}) are replaced by the
respective averages of their values on the corner vertices.  The
partial derivatives occuring in (\ref{energydens2}) follow from the
linear approximations (\ref{interpol1}) and (\ref{interpol2}).

\newpage


\newpage

%
%
\begin{figure}
\caption{Definition of the toroidal geometry. (a) Section across
$x$-$z$ plane (side view), (b) section across $x$-$y$ plane (top view).
The radii $A$ and $B$ denote the geometrical parameters of the
confining torus.}
\end{figure}

%
%
\begin{figure}
\caption{Fixed director field at the torus surface
in sections across the $x$-$z$ and $y$-$z$ coordinate planes.
(a) Homeotropic, (b) transversal-tangential, (c) azimuthal-tangential
surface anchoring.}
\end{figure}

%
%
\begin{figure}
\caption{Equilibrium director field in sections
across the $x$-$z$ and $y$-$z$ coordinate planes
for homeotropic surface anchoring.
(a) One defect ring of topological strength $+1$,
(b) two defect rings of topological strength $+\frac{1}{2}$,
(c) director ``escape'' along azimuthal direction.}
\end{figure}

%
%
\begin{figure}
\caption{Total energy $\overline{F}$ per volume
of different director configurations
vs.~inverse magnetic coherence length $\overline{\xi}$, for
homeotropic surface anchoring. 
Squares:
one-ring configuration, homogeneous magnetic field along $z$.
$\times$-symbols:
two-ring configuration, homogeneous magnetic field along $z$.
Rhombs: escape configuration, azimuthal magnetic field.
The upper and lower plots refer to the torus radii $A = 7$, $B= 2$
and $A = 7$, $B = 2.4$, respectively.}
\end{figure}

%
%
\begin{figure}
\caption{Equilibrium director field in sections
across the $x$-$z$ and $y$-$z$ coordinate planes
for transversal-tangential surface anchoring.
(a) One defect ring of topological strength $+1$ (vortex),
(b) two surface defect rings of topological strength $+\frac{1}{2}$,
(c) director ``escape'' along azimuthal direction.}
\end{figure}

%
%
\begin{figure}
\caption{Total energy $\overline{F}$ per volume
of different director configurations
vs.~inverse magnetic coherence length $\overline{\xi}$, for
transversal-tangential surface anchoring. 
Squares:
one-ring configuration (vortex), homogeneous magnetic field along $z$.
$\times$-symbols:
two-ring configuration, homogeneous magnetic field along $z$.
Rhombs: escape configuration, azimuthal magnetic field.
The upper and lower plots refer to the torus radii $A = 7$, $B= 2$
and $A = 7$, $B = 2.4$, respectively.}
\end{figure}

%
%
\begin{figure}
\caption{Equilibrium director field in sections
across the $x$-$z$ and $y$-$z$ coordinate planes
for azimuthal-tangential surface anchoring.
(a) Azimuthal configuration below the magnetic threshold field,
(b) distorted configuration above the magnetic threshold field.}
\end{figure}

%
%
\begin{figure}
\caption{Total energy $\overline{F}$ per volume
of the director field subject to
a homogeneous magnetic field along $z$
vs.~inverse magnetic coherence length $\overline{\xi}$, for
azimuthal-tangential surface anchoring. 
Squares: torus radii $A = 7$, $B = 2$.
$\times$-symbols: torus radii $A = 7$, $B = 2.4$.}
\end{figure}

%
%
\begin{figure}
\caption{Equilibrium director field subject to a lateral homogeneous
magnetic field along $x$ with inverse coherence length 
$\overline{\xi} = 2.5$, for homeotropic surface anchoring.
(a) Section across the $x$-$y$ plane (top view),
(b) sections
across the $x$-$z$ and $y$-$z$ coordinate planes (oblique view).}
\end{figure}

%
%
\begin{figure}
\caption{Equilibrium director field subject to a lateral homogeneous
magnetic field along $x$ with inverse coherence length 
$\overline{\xi} = 2.5$, for transversal-tangential surface anchoring.
(a) Section across the $x$-$y$ plane (top view),
(b) sections
across the $x$-$z$ and $y$-$z$ coordinate planes (oblique view).}
\end{figure}

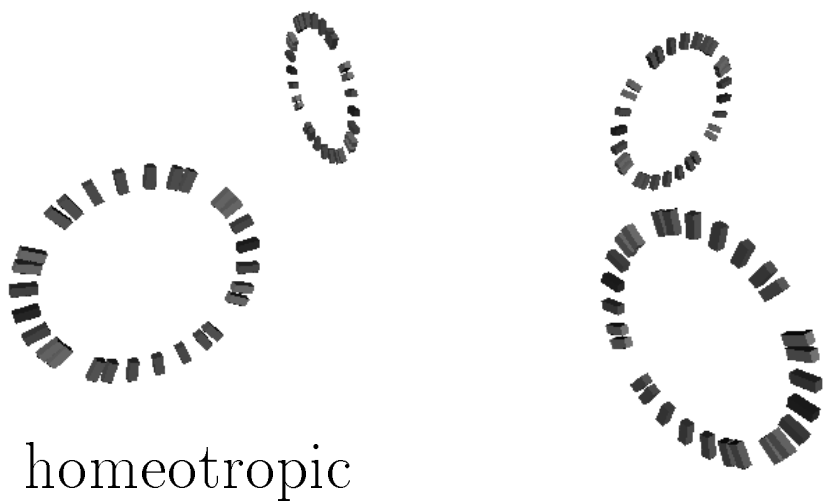
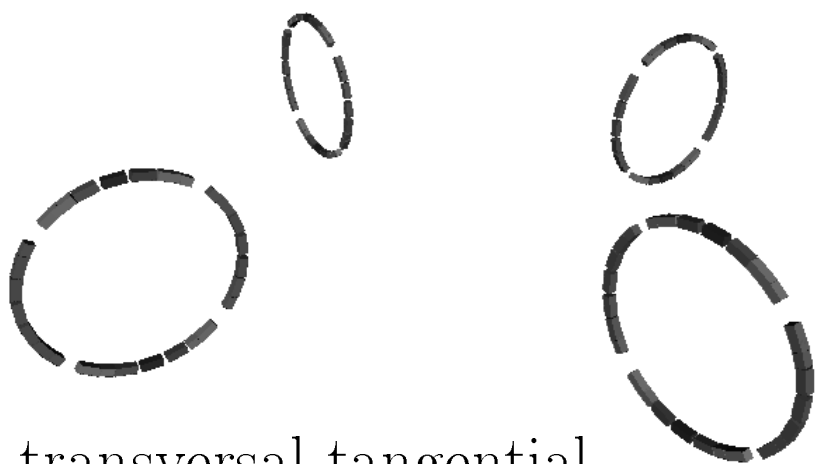
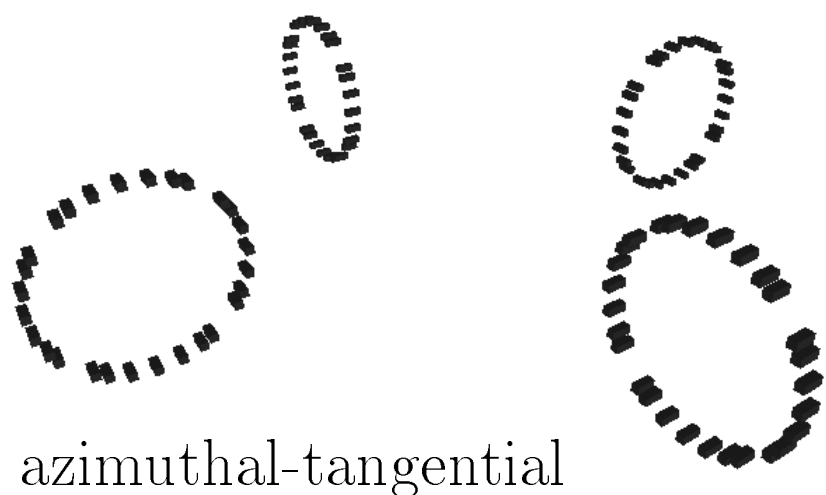
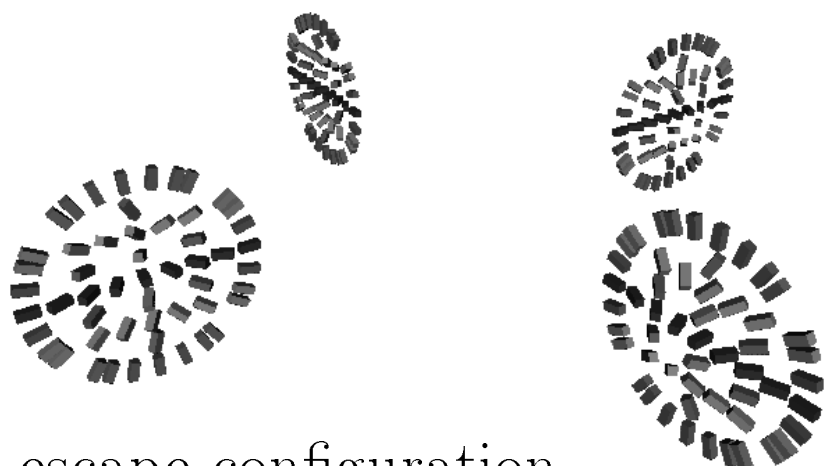
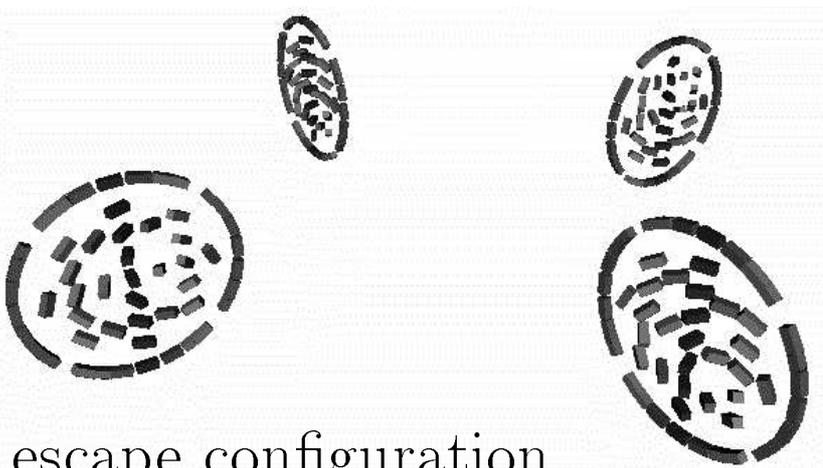
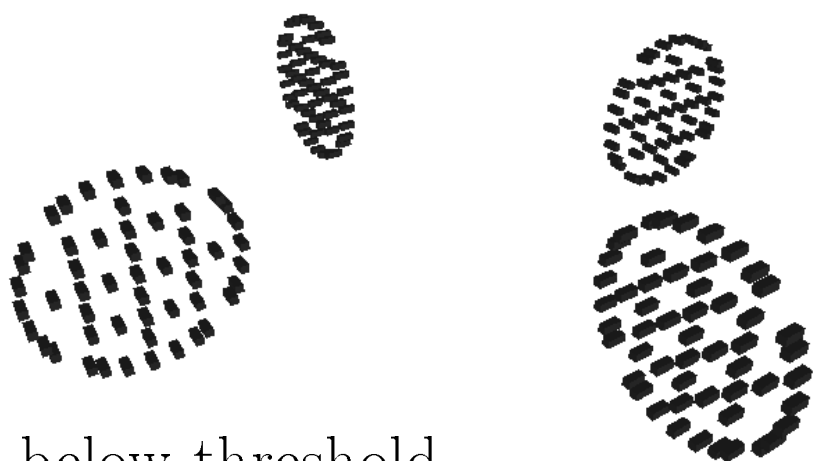
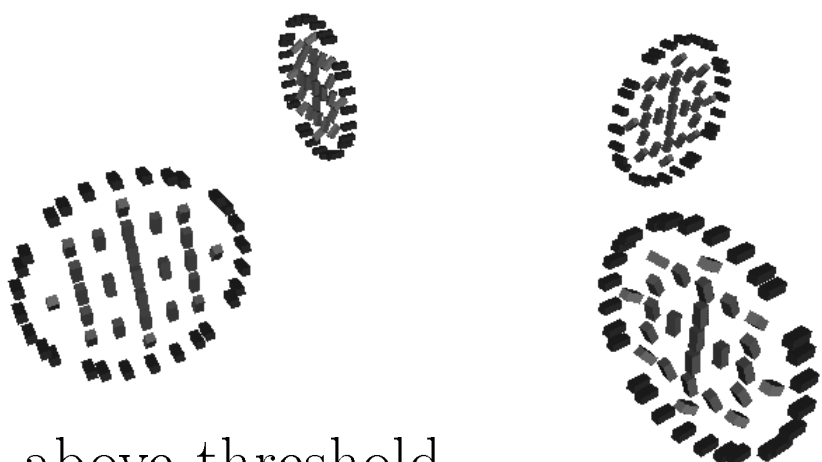
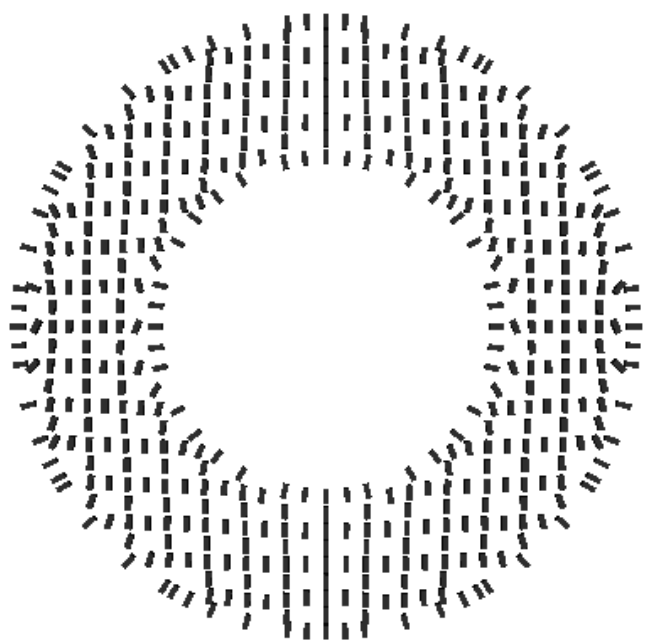
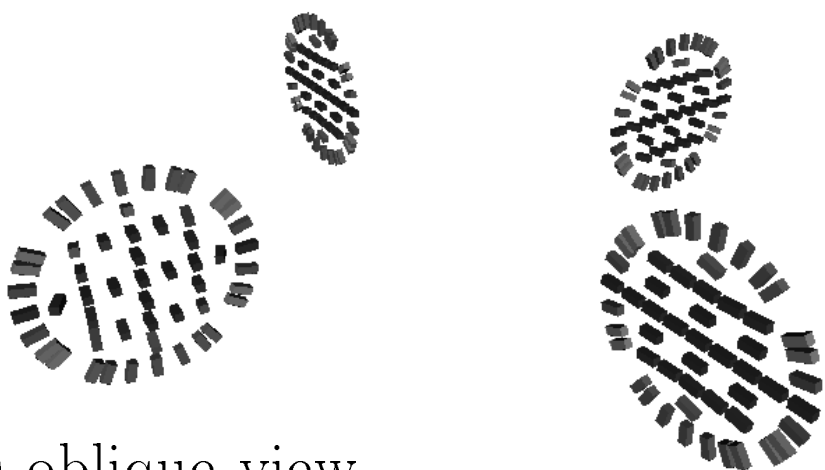
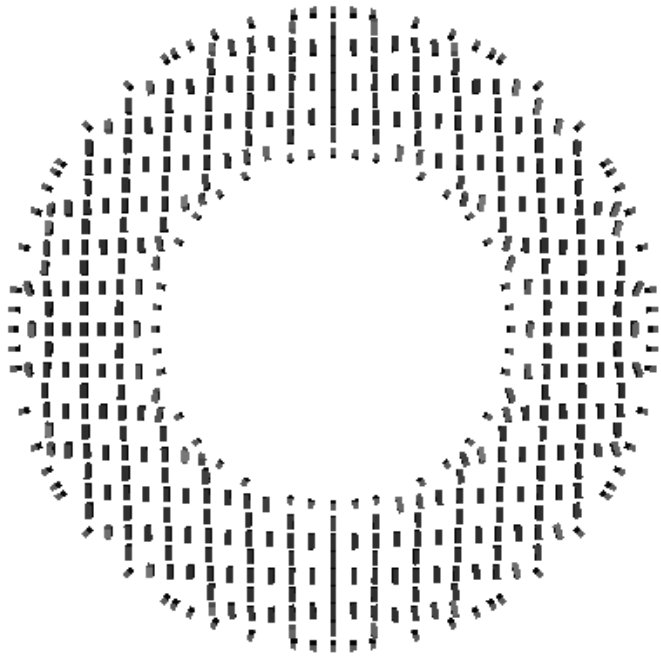
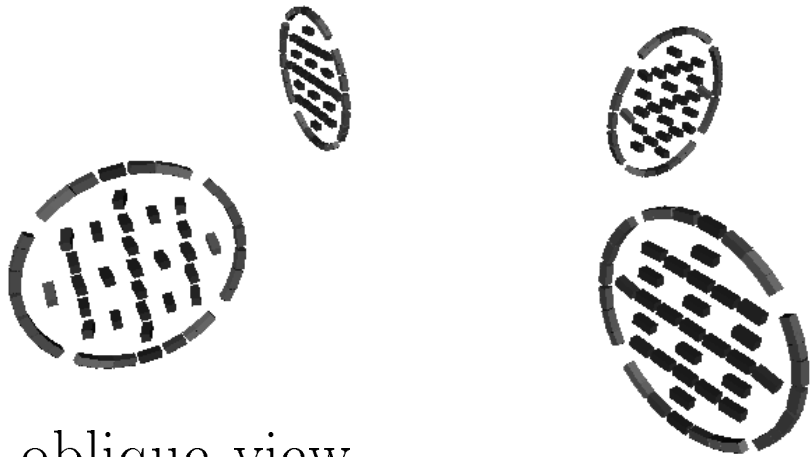
\end{document}